\documentstyle[amssymb,aps,graphicx]{revtex}
 \def\ket{\rangle}
\def\<{\langle}
\def\>{\rangle}

\begin{document}
\title{ Cooperative three- and four- player quantum games\thanks{To appear in Phys. Lett. A,
Correspondence to: gllong@tsinghua.edu.cn}}
\author{Ying Jun Ma$^{1,2}$, Gui Lu
Long$^{1,2,3,4}$, Fu Guo Deng$^{1,2}$, Feng Li$^{5}$ and
Sheng-Xing Zhang$^{1,2}$}
\address{
 $^1$Department of Physics, Tsinghua University, Beijing 100084,
China\\
$^2$Key Laboratory For Quantum Information and Measurements,
Beijing 100084, China\\
$^3$ Center for Atomic and Molecular NanoSciences, Tsinghua
University, Beijing 100084, China\\
$^4$Institute of Theoretical Physics, Chinese Academy of
Sciences, Beijing 100080, China\\
$^5$ Basic Education Section, Capital University of Economics and
Business, West Campus, Beijing 100070, China}

\maketitle

\begin{abstract}

A cooperative multi-player quantum game played by 3 and 4 players
has been studied. Quantum superposed operator is introduced in
this work which solves the non-zero sum difficulty in previous
treatment. The role of quantum entanglement of the initial state
is discussed in details.
\end{abstract}


\section{Introduction}
\label{s1}

 Quantum game theory has become a new area of application of quantum theory. It has
attracted much attention. Among them, two-player quantum game
\cite{r1,r2}, multi-player non-cooperative quantum game \cite{r3}
and the cooperative three player quantum game \cite{r4} have been
reported recently. The prisoner's dilemma game has been
demonstrated recently in NMR experiment\cite{r4p}. In classical
game theory, the cooperation of the players means that they can
exchange completely information with one another, and they take
the strategy to the most behoof of themselves. In this way,
co-operators can be seen as one player. It is very important in
cooperative game that each party in a cooperation must take
coordinated strategies. For this purpose, in game theory, Nash
equilibrium(NE) is an important concept. In an Nash equilibrium,
each player obtains his/her optimal payoff, and if he/she tries to
change his/her strategy from the NE strategy, his/her payoff will
become less. In cooperative quantum game, the situation is
similar.

In this paper, we study a  game played by 3 and 4 players. The
rules for the cooperative 3 player game were given in
Ref.\cite{r4}.  Suppose that there are  three players denoted as
$N=\{A,B,C\}$,  where $A$ and $B$
are cooperators are  $C$ is solitary. Each player has a strategy set $%
S_i(i=A,B,C)$. The payoff set for the corresponding strategies is
$P_i(i=A,B,C)$. It is a function of the strategy $S_A\times
S_B\times S_C$. Each player can choose one of the two strategies
denoted as $0$ and $1$. In each play, if there is a player uses
the minority strategy, then this player is a looser and he/she has
to give one penny to each of the other two players. In the 4
player cooperative  game,  the model and the rules are similar.
But in this case, there is a possibility that every two players
choose the same strategy, then there is no gain or loss for
anyone.

In the quantum version of this game, an multi-qubit initial
quantum state is prepared by an arbiter and the qubits are sent
to all players at random, one qubit for each player.
 It is assumed that each player possesses local
unitary operators $i.e.$ $I$ and $\sigma _x$ on their own qubit
at their proposal, and then sends the qubit back to the arbiter.
Upon receiving all the qubits, the arbiter will measure them to
get the results and distribute the payoff to each player. In this
work, we studied two different ways each player performs his/her
strategies. In the classical probability operator(CPO) way, each
player is allowed to perform $I$ and $\sigma_x$ with probability
$p$ and $1-p$ respectively. In CPO, the players work exactly the
way as in a classical game. The only difference is that they lay
their bet on a quantum state. In Ref.\cite{r4,r5,r6}, the
strategy is performed in this way. In a quantum superposed
operator(QSO) way, each player is allowed to perform an unitary
operation that is a linear combination of $I$ and $\sigma_x$:
$\sqrt{p}I+\sqrt{1-p}\sigma_x$. This corresponds to a physical
implementation that each player is allowed to use linear
superposition of the identity operator and the $\sigma_x$. Acting
upon state $|0\ket$, it produces
$\sqrt{p}|0\ket+\sqrt{1-p}|1\ket$. If one makes measurement on
this state, he/she will have $p$ probability to obtain 0, and
$1-p$ probability to obtain 1. This is closely related  to  CPO
and classical game with $p$ probability to choose 0 and $1-p$
probability to choose 1.
 The cooperative quantum game
between 3 players using CPO has been studied in Ref.\cite{r4}. In
our work, we study the 3 and 4 player quantum minority game using
QSO and the 4 player game using CPO. We found that qualitatively,
QSO and CPO produce the same outcome, but they differ in details
in quantity. It is found that  in a cooperative quantum game, the
initial state plays a crucial role. For some initial state,
cooperation is very important. For some other initial state,
cooperation becomes useless. This is quite different from what we
know in a classical game theory.

\section{Three-player  cooperative game in QSO}
\label{s2}

Now we come to the three cooperative quantum game (TCQG). The players
implement their strategies by applying the identity operator $I$ and the $%
Pauli$-spin flip $\sigma _x.$ The three player cooperative game
with CPO has been studied in Ref. \cite{r4}. Here we will discuss
this game with QSO. We assume that the cooperators take the
correlated operations to the their qubits. Now suppose the initial state is $%
|\psi _0\ket$=$c_1|000\ket$+$c_2|111\ket$, where $c_1^2+c_2^2=1$, $c_1\geq 0$, $%
c_2\geq 0$. Because $A$ and $B$ are cooperators, in term of the
rules, they take the same strategies, so they do the same
operations on their respective qubits:
 $\sqrt{p}II+\sqrt{1-p}%
\sigma _x\sigma _x$. $C\ \ $takes operation $\sqrt{q}I+\sqrt{1-q}%
\sigma _x$. After the operations have been performed, the state
of the system becomes
\begin{eqnarray}
\psi _{out} &= &(c_1\sqrt{pq}+c_2\sqrt{(1-p)(1-q)})|000\ket \nonumber \\
&&+(c_1\sqrt{p(1-q)}+c_2\sqrt{(1-p)q})|001\ket  \nonumber \\
&&+(c_1\sqrt{(1-p)q}+c_2\sqrt{p(1-q)})|110\ket  \nonumber \\
&&+(c_1\sqrt{(1-p)(1-q)}+c_2\sqrt{pq})|111\ket. \label{e1}
\end{eqnarray}

The payoff of player A is the sum of squares of the coefficients
of $|001\ket$ and $|110\ket$:
\begin{equation}
P_A=p+q-2pq+4c_1c_2\sqrt{p(1-p)q(1-q)}=p+q-2pq+4a\sqrt{p(1-p)q(1-q)},
\label{e2}
\end{equation}
where $a=c_1c_2$. We have plotted the payoff of A with respect to
$p$ and $q$ in Fig.1. ``$a$'' is a measure of entanglement of the
initial state. When ``$a$''=0, the initial state is a product
state and there is no entanglement in it. We see from Fig.1a that
there is clearly a saddle point in the 3-dimensional figure. As
``$a$'' increases to $1/3$, the saddle point becomes
shallower(Fig.1b). When ``$a$''=1, the system has maximum
entanglement, and the saddle becomes flat(Fig.1c). This clearly
shows that the initial state is very important in the cooperative
game.  The 3D-figure displays a shape of saddle, and the
saddle-point is just the NE point. To gain the optimal payoff, we
let
\begin{eqnarray}
\frac{\partial P_A}{\partial p}
&=&1-2q+2a\sqrt{\frac{q(1-q)}{p(1-p)}}(1-2p)=0,
\label{e3a} \\
\frac{\partial P_A}{\partial q}
&=&1-2p+2a\sqrt{\frac{p(1-p)}{q(1-q)}}(1-2q)=0. \label{e3b}
\end{eqnarray}
\ \

Combining Eqs.\label{e3 copy(1)}(\ref{e3a}, \ref{e3b}), we can get $%
p=q=1/2$. Substituting into Eq.(\ref{e2})\label{e2 copy(2)}, then
the payoff $A$ can get is
\begin{equation}
P_{A,max}=1/2+a.  \label{e4}
\end{equation}

The relation between $P_{A,max}$ and $a$ is shown in Fig. 2.
Similarly, the payoff of $B$ and $C$ are:
\begin{eqnarray}
P_B &=&P_A=1/2+a,   \label{e5} \\
P_C &=&-2P_A=-1-2a.
\end{eqnarray}

For players $A$ and $B$, the larger the numerical value of $a$,
the more their payoff. The payoff of player $A$ varies
continuously with the change in the value of $a$. When $a=0$, in
other words, the initial state is $|000\ket$ or $|111\ket$, the
payoff is the same as  that in a  classical game, $viz$ the
payoff is 0.5. When the initial state has maximum entanglement,
his payoff achieves the
maximum, 1. Thus,  cooperation must have more advantage for $A$ and $%
B$, and their cooperation will be all to nothing going on. Certainly, $%
P_A+P_B+P_C=0$, that is, our scheme is still a zero-sum game but that in
Ref. \cite{r4} is not.

In fact, when the initial state is $c_1|001\ket$+$c_2|110\ket$,
we can use the same method to find out the equilibrium point.
Similarly, the players can take the strategies described above.
Then,
\begin{eqnarray}
\psi _{out} &= &\ \quad
(c_1\sqrt{p(1-q)}+c_2\sqrt{(1-p)q})|000\ket
\nonumber  \\
&&+(c_1\sqrt{(1-p)(1-q)}+c_2\sqrt{pq})|110\ket  \nonumber \\
&&+(c_1\sqrt{(1-p)q}+c_2\sqrt{p(1-q)})|111\ket. \label{e6}
\end{eqnarray}
In term of this equation, we can write the payoff of $A$:
\begin{equation}
P_A=1-p-q+2pq+4a\sqrt{p(1-p)q(1-q)}.  \label{e7}
\end{equation}
The relation between $P_A$ and these parameters can be seen from
Fig. 3. The saddle-point still exists. We can find out the
maximum payoff  by solving the following equations:
\begin{eqnarray}
\frac{\partial P_A}{\partial p} &=&-1+2q+2a\frac{q(1-q)(1-2p)}{\sqrt{%
p(1-p)q(1-q)}},  \nonumber   \\
\frac{\partial P_A}{\partial q} &=&-1+2p+2a\frac{p(1-p)(1-2q)}{\sqrt{%
p(1-p)q(1-q)}}.\label{e8}
\end{eqnarray}
We get the saddle-point at $p=1/2$ and $q=1/2$, and then $P_A=1/2+a$, $%
P_B=1/2+a$, $P_C=-1-2a$. The result is the same as that discussed
just above.  In the classical game, the NE strategy is a point,
and so is their corresponding payoff. But in the quantum game,
their NE strategies depend on the initial entangle state.
Furthermore, in this situation, the cooperation must be of
advantage for players $A$ and $B$, and this is the very reason
for their continuance of cooperation.

\section{Four-player cooperative quantum game in CPO}
\label{s3}

Up to now, no one has studied the four cooperative quantum game
(FCQG). In this section we will investigate FCQG in CPO. Here
what the players do is similar to that metioned above. The
initial state $|\psi _{in}\rangle
=\sum_{i,j,k,l=0}^1{C}_{ijkl}{|}ijkl{\rangle }$, where $\sum_{i,j,k,l=0}^1%
\left| C_{ijkl}\right| ^2=1$. Each player takes operation $I$ with classical
probabilities $p${\it ,}$q${\it , }$r${\it , }$l$, respectively, but the $%
Pauli$-spin-flip operator $\sigma _x$ with probabilities $1-p$, $1-q $, $1-r$%
, $1-l,$ respectively. After the performance of the strategies,
the state of the system changes from $|\psi _{in}\rangle $ to an
mixed state represented by a density operator, $\rho
=\sum_{i,j,k,l=0}^1{|}P_{ijkl}{|^2|}ijkl{\rangle \langle
}ijkl{|}$. In fact, the state becomes an mixed state when player
take strategy in CPO manner.  For convenience, the following
notations
are introduced: $P_0=|P_{0000}|^2$, $P_1=|P_{0001}|^2$, $\cdots $, $%
P_{15}=|P_{1111}|^2$,  which are the probabilities of the system
ending up at state $[0000]$, $[0001]$, $...$, $[1111]$
respectively.  $x_0=|C_{0000}|^2$, $x_1=|C_{0001}|^2$, $\cdots $,
$x_{15}=|C_{1111}|^2$ are the probabilities of the system being
in state $|0000\ket$, $|0001\ket$, $...$, $|1111\ket$
respectively. The relation between these quantities is
\begin{equation}
Q=\Lambda \;\Pi ^T,  \label{e9}
\end{equation}
where
\begin{equation}
Q=\left[ P_0,P_1,\cdots ,P_{15}\right] ^T  \label{e10}
\end{equation}
and

\begin{eqnarray}
\Pi  &=&[pqrl,pqr\bar{l},pq\bar{r}l,
pq\bar{r}\bar{l},p\bar{q}rl,p\bar{q}r\bar{l},
p\bar{q}\bar{r}l,p\bar{q}\bar{r}\bar{l}, \bar{p}qrl,
\bar{p}qr\bar{l},\bar{p}q\bar{r}l,\bar{p}q\bar{r}\bar{l},
\bar{p}\bar{q}rl,\bar{p}\bar{q}r\bar{l},\bar{p}\bar{q}\bar{r}l,
\bar{p}\bar{q}\bar{r}\bar{l}],
\end{eqnarray}
where a bar over a character means 1 minus this quantity, e.g.
$\bar{p}=1-p$.
 The matrix $\Lambda $ is calculated as follows
\begin{equation}
\Lambda =\left[
\begin{array}{cccccccccccccccc}
x_0 & x_1 & x_2 & x_3 & x_4 & x_5 & x_6 & x_7 & x_8 & x_9 & x_{10} & x_{11}
& x_{12} & x_{13} & x_{14} & x_{15} \\
x_1 & x_0 & x_3 & x_2 & x_5 & x_4 & x_7 & x_6 & x_9 & x_8 & x_{11} & x_{10}
& x_{13} & x_{12} & x_{15} & x_{14} \\
x_2 & x_3 & x_0 & x_1 & x_6 & x_7 & x_4 & x_5 & x_{10} & x_{11} & x_8 & x_9
& x_{14} & x_{15} & x_{12} & x_{13} \\
x_3 & x_2 & x_1 & x_0 & x_7 & x_6 & x_5 & x_4 & x_{11} & x_{10} & x_9 & x_8
& x_{15} & x_{14} & x_{13} & x_{12} \\
x_4 & x_5 & x_6 & x_7 & x_0 & x_1 & x_2 & x_3 & x_{12} & x_{13} & x_{14} &
x_{15} & x_8 & x_9 & x_{10} & x_{11} \\
x_5 & x_4 & x_7 & x_6 & x_1 & x_0 & x_3 & x_2 & x_{13} & x_{12} & x_{15} &
x_{14} & x_9 & x_8 & x_{11} & x_{10} \\
x_6 & x_7 & x_4 & x_5 & x_2 & x_3 & x_0 & x_1 & x_{14} & x_{15} & x_{12} &
x_{13} & x_{10} & x_{11} & x_8 & x_9 \\
x_7 & x_6 & x_5 & x_4 & x_3 & x_2 & x_1 & x_0 & x_{15} & x_{14} & x_{13} &
x_{12} & x_{11} & x_{10} & x_9 & x_8 \\
x_8 & x_9 & x_{10} & x_{11} & x_{12} & x_{13} & x_{14} & x_{15} & x_0 & x_1
& x_2 & x_3 & x_4 & x_5 & x_6 & x_7 \\
x_9 & x_8 & x_{11} & x_{10} & x_{13} & x_{12} & x_{15} & x_{14} & x_1 & x_0
& x_3 & x_2 & x_5 & x_4 & x_7 & x_6 \\
x_{10} & x_{11} & x_8 & x_9 & x_{14} & x_{15} & x_{12} & x_{13} & x_2 & x_3
& x_0 & x_1 & x_6 & x_7 & x_4 & x_5 \\
x_{11} & x_{10} & x_9 & x_8 & x_{15} & x_{14} & x_{13} & x_{12} & x_3 & x_2
& x_1 & x_0 & x_7 & x_6 & x_5 & x_4 \\
x_{12} & x_{13} & x_{14} & x_{15} & x_8 & x_9 & x_{10} & x_{11} & x_4 & x_5
& x_6 & x_7 & x_0 & x_1 & x_2 & x_3 \\
x_{13} & x_{12} & x_{15} & x_{14} & x_9 & x_8 & x_{11} & x_{10} & x_5 & x_4
& x_7 & x_6 & x_1 & x_0 & x_3 & x_2 \\
x_{14} & x_{15} & x_{12} & x_{13} & x_{10} & x_{11} & x_8 & x_9 & x_6 & x_7
& x_4 & x_5 & x_2 & x_3 & x_0 & x_1 \\
x_{15} & x_{14} & x_{13} & x_{12} & x_{11} & x_{10} & x_9 & x_8 & x_7 & x_6
& x_5 & x_4 & x_3 & x_2 & x_1 & x_0
\end{array}
\right] .
\end{equation}
The matrix is symmetric. The payoff of the players $A$, $B$, $C$,
$D$ in each basis state $|i\rangle $ are denoted as $\alpha _i$,
$\beta _i$, $\gamma _i$, $\theta _i$, where range of $i$ is from
$0$ through $15$. For instance for player $A$, the payoff in the
basis states are
\begin{eqnarray}
&&\alpha _0=\alpha _3=\alpha _5=\alpha _6=\alpha _9=\alpha _{10}=\alpha
_{12}=\alpha _5=0,  \label{e12} \nonumber\\
&&\alpha _1=\alpha _2=\alpha _4=\alpha _{11}=\alpha _{13}=\alpha
_{14}=1, \alpha _7=\alpha _8=-3.
\end{eqnarray}
In the end of the game, the payoff function of player $A$ is
\begin{eqnarray}
P_A &=&\sum_{i=0}^{15}P_i\alpha _i  \nonumber  \\
&=&(x_0+x_{15})(2pr+2pl+2qp-2ql-2rl-2qr+l+r+q-3p)  \nonumber \\
&&+(x_1+x_{14})(2pq+2pr+2ql-2pl-2qr+2rl-p-q-l-r+1)  \nonumber \\
&&+(x_2+x_{13})(2pq+2qr+2pl+2rl-2pr-2ql-p-q-l-r+1)  \nonumber \\
&&+(x_3+x_{12})(2qr+2ql+2pq-2pr-2pl-2rl-3q+r+p+l)  \nonumber \\
&&+(x_4+x_{11})(2pr+2pl+2qr+2ql-2qp-2rl-p-q-r-l+1)  \nonumber \\
&&+(x_5+x_{10})(2pr+2qr+2rl-2qp-2ql-2pl+l+p+q-3r)  \nonumber \\
&&+(x_6+x_9)(2pl+2ql+2rl-2qp-2qr-2pr+q+r+p-3l)  \nonumber \\
&&+(x_7+x_8)(-2pq-2ql-2rl-2pr-2pl-2qr+3p+3q+3r+3l-3). \label{e13}
\end{eqnarray}
For cooperative game, we let
\begin{eqnarray}
&&x_4=x_{11}= x_5=x_{10}=x_6=x_9=x_7=x_8=0.\label{e14}
\end{eqnarray}
Then the corresponding payoff for player A will be simplified
from eq.(\ref{e13}) with those terms deleted. Under this
condition, the payoff for player $C$ is
\begin{eqnarray}
&&P_{C,co}=(x_0+x_{15})(2pr+2qr+2rl-2qp-2pl-2ql+p+q+l-3r)  \nonumber \\
&&+(x_1+x_{14})(2pr+2pl+2qr+2ql-2qp-2rl-p-q-r-l+1)  \nonumber \\
&&+(x_2+x_{13})(-2qp-2pr-2qr-2pl-2ql-2rl+3r+3p+3q+3l-3)  \nonumber \\
&&+(x_3+x_{12})(2pl+2ql+2rl-2qp-2pr-2qr+p+q+r-3l).  \label{e16}
\end{eqnarray}

Initial state in quantum game is important in deciding the fairness of the game.
Next we discuss some special initial states. First if the initial state has the form where $%
x_0=x_1=x_2=x_3=x_{12}=x_{13}=x_{14}=x_{15}=1/8$, then we know
$P_A=P_B$ and $P_C=P_D$, and
\begin{eqnarray}
&&P_A =P_B=\frac 14(8qp-4p-4q+2),   \nonumber \\
&&P_C =P_D=\frac 14(-8qp+4p+4q-2),  \nonumber \\
&&P_A+P_B+P_C+P_D =0. \label{e17}
\end{eqnarray}
In general, quantum game in CPO is not zero-sum. This is a
zero-sum game is obtained.
For this initial state, players $A$ and $B$ will choose $p=q=0$ or $%
p=q=1$ to gain the maximum payoff. They get
$P_{Amax}=P_{Bmax}=1/2$. Although players $A$ and $B$ are
dominate, their maximum payoff is not better than that of
classical counterpart.

In another example with $x_0=1$, no entanglement exists in the
initial state, we have
\begin{eqnarray}
P_A &=&2pr+2pl+2qp-2ql-2rl-2qr+l+r+q-3p,  \label{e18} \\
P_C &=&2pr+2qr+2rl-2qp-2pl-2ql+p+q+l-3r.
\end{eqnarray}
When $p=q$, both the payoff of $A$ and $B$ are $2q^2-2rl-2q+r+l$.
Equilibrium point is obtained at $p=q=1$ or $p=q=0$ together with
$r=l=1/2$. The payoffs for players A, B, C and D are $1/2 $,
$1/2$, $-1/2$, $-1/2$ respectively. The sum of the payoff of four
players $P_A+P_B+P_C+P_D=0$ is also zero-sum.

The sensitive dependence of the results on initial state is best
explained by the following example where we allow the initial
state to have some components of non-cooperation, such as
components with nonzero $x_5$, $x_{10}$, $x_6$ and $X_9$.
Assuming an initial state with
\begin{eqnarray}
x_1 =x_{14}=0, \;x_2 =x_{13}=0,\;x_4 =x_{11}=0,\; x_7 =x_8=0,
\end{eqnarray}
the payoffs of player A is then
\begin{eqnarray}
&&P_A=(x_0+x_{15})(2pr+2qp+2pl-2ql-2rl-2qr+l+r+q-3p)  \nonumber \\
&&+(x_3+x_{12})(2qr+2ql+2qp-2pr-2pl-2rl-3q+r+p+l)  \nonumber \\
&&+(x_5+x_{10})(2pr+2qr+2rl-2qp-2ql-2pl+l+p+q-3r)  \nonumber \\
&&+(x_6+x_9)(2pl+2ql+2rl-2qp-2qr-2pr+q+r+p-3l).  \label{PA}
\end{eqnarray}
Expression for  $P_B$ is obtained by exchanging $p$ and $q$ in
Eq.(\ref{PA}). The payoff for player C is
\begin{eqnarray}
&&P_C=(x_0+x_{15})(2pr+2qr+2rl-2qp-2pl-2ql+p+q+l-3r)  \nonumber \\
&&+(x_3+x_{12})(2pl+2ql+2rl-2qp-2pr-2qr+p+q+r-3l)  \nonumber \\
&&+(x_5+x_{10})(2pr+2pl+2qp-2qr-2rl-2ql+r+q+l-3p)  \nonumber \\
&&+(x_6+x_9)(2qr+2ql+2qp-2pr-2pl-2rl+p+r+l-3q).  \label{PC}
\end{eqnarray}
By exchanging $r$ with $l$ in Eq.(\ref{PC}), expression for $P_D$
can be obtained. If the coefficients is chosen as
$x_0=x_3=x_5=x_6=x_9=x_{10}=x_{12}=x_{15}=1/8$, then we will
arrive at an equilibrium. All the payoffs become zero. Under this
circumstance, whatever strategies the players take, their payoff
are all zero. It is the same as a complete non-cooperative quantum
game, and the advantages of the cooperators in a classical game
are completely eliminated.

\section{Quantum 4 player cooperative game in QSO}
\label{s4}

 Suppose the initial state is $|\psi _0\ket=c_1|0000\ket+c_2|1111\ket$.
Players $A$ and $B$ take their correlated operation
$\sqrt{p}II+\sqrt{1-p}\sigma _x\sigma _x$, player
$C$ and $D$  take strategy $\sqrt{q}I+\sqrt{1-q}\sigma _x$ and $\sqrt{l%
}I+\sqrt{1-l}\sigma _x$ respectively. After the operations of
their strategies, the state of the system becomes
\begin{eqnarray}
\psi _{out}&=& (c_1\sqrt{pql}+c_2\sqrt{(1-p)(1-q)(1-l)}%
)|0000\ket  \nonumber   \\
&&+(c_1\sqrt{pq(1-l)}+c_2\sqrt{(1-p)(1-q)l})|0001\ket  \nonumber \\
&&+(c_1\sqrt{p(1-q)l}+c_2\sqrt{(1-p)q(1-l)})|0010\ket  \nonumber \\
&&+(c_1\sqrt{p(1-q)(1-l)}+c_2\sqrt{(1-p)ql})|0011\ket  \nonumber \\
&&+(c_1\sqrt{(1-p)ql}+c_2\sqrt{p(1-q)(1-l)})|1100\ket  \nonumber \\
&&+(c_1\sqrt{(1-p)q(1-l)}+c_2\sqrt{p(1-q)l})|1101\ket  \nonumber \\
&&+(c_1\sqrt{pq(1-l)}+c_2\sqrt{(1-p)(1-q)l})|1110\ket  \nonumber \\
&&+(c_1\sqrt{(1-p)(1-q)(1-l)}+c_2\sqrt{pql})|1111\ket.\label{e22}
\end{eqnarray}

By summing up of square of the related coefficients, $P_A$ is
obtained as follows
\begin{equation}
P_A=q+l+8a\sqrt{p(1-p)q(1-q)l(1-l)}-2lq,  \label{e23}
\end{equation}
where $a=c_1c_2$. The equilibrium can be obtained by solving the
following equations
\begin{eqnarray}
&&\frac{\partial P_A}{\partial p}=4a\frac{(1-p)q(1-q)l(1-l)-pq(1-q)l(1-l)}{%
\sqrt{p(1-p)q(1-q)l(1-l)}}=0,  \label{e24a} \\
&&\frac{\partial P_A}{\partial q}=1+4a\frac{p(1-p)(1-q)l(1-l)-pq(1-p)l(1-l)}{%
\sqrt{p(1-p)q(1-q)l(1-l)}}-2l=0,  \label{e24b} \\
&&\frac{\partial P_A}{\partial l}=1+4a\frac{p(1-p)q(1-q)(1-l)-pq(1-p)(1-q)l}{%
\sqrt{p(1-p)q(1-q)l(1-l)}}-2q=0.  \label{e24c}
\end{eqnarray}

The solution is $p=1/2$, $q=1/2$, $l=1/2$. This result doesn't
depend on the initial state parameters $c_1$ and $c_2$.
Substituting this to Eq. (\ref{e23}), the payoff is $P_A=1/2+a$.
 the payoff depends on the initial state  entanglement parameter $a$.
The dependence is exactly the same as that shown in Fig.2.
Compared with the classical result, there is additional parameter
$a$ and it plays a key role
in determining the value of payoff. Furthermore, once $a=0$, the payoff $%
P_A=1/2$ and the quantum properties of the game disappeared,
namely, the game degenerated to  a classical one. Meanwhile, the
payoffs of $B$, $C$ and $D$ are:
\begin{eqnarray}
P_B &=&P_A,  \label{e25} \\
P_c &=&-8c\sqrt{p(1-p)q(1-q)l(1-l)}+4pq-4lp+2lq-3q+l, \\
P_D &=&-8c\sqrt{p(1-p)q(1-q)l(1-l)}-4pq+4lp+2lq-q-3l.
\end{eqnarray}
We can verify easily that $P_A+P_B+P_C+P_D=0$, it is a zero-sum
game. For the maximum entanglement state, the corresponding
payoff of $A$ takes its maximum $1$. It is twice as much as that
in a  classical game.

Next we consider another initial state,
$c_1|0001\ket+c_2|1101\ket$. After the completion of strategy
operations, the state of the system becomes:
\begin{eqnarray}
\psi _{out} &=& (c_1\sqrt{pq(1-l)}+c_2\sqrt{(1-p)q(1-l)}%
)|0000\ket  \nonumber   \\
&&+(c_1\sqrt{pql}+c_2\sqrt{(1-p)ql})|0001\ket  \nonumber \\
&&+(c_1\sqrt{p(1-q)(1-l)}+c_2\sqrt{(1-p)(1-q)(1-l)})|0010\ket  \nonumber \\
&&+(c_1\sqrt{p(1-q)l}+c_2\sqrt{(1-p)(1-q)l})|0011\ket  \nonumber \\
&&+(c_1\sqrt{(1-p)q(1-l)}+c_2\sqrt{pq(1-l)})|1100\ket  \nonumber \\
&&+(c_1\sqrt{(1-p)ql}+c_2\sqrt{pql})|1101\ket  \nonumber \\
&&+(c_1\sqrt{(1-p)(1-q)(1-l)}+c_2\sqrt{p(1-q)(1-l)})|1110\ket  \nonumber \\
&&+(c_1\sqrt{(1-p)(1-q)l}+c_2\sqrt{p(1-q)l})|1111\ket.\label{e26}
\end{eqnarray}

We find that the payoff of player $A$ is $(1-q-r+2qr)(1+4a\sqrt{p(1-p)}$, and $%
p=1/2$, $q=1/2$, $l=1/2$ is the NE point where $P_A=a+1/2$. This
result is the same as that mentioned above.

\section{Conclusion}

\label{s5}

We have studied the three- and four- player cooperative quantum
games. The expressions for the functions of payoff of the players
are given. It is found that in a cooperative quantum game, the
initial state is important for all players. For some special
initial states, the cooperation will become  important.  When the
initial state is direct-product state, cooperative game becomes
classical game. The GSO strategy is introduced which has a close
correspondence with classical game. In particular, it preserves
the zero-sum property in classical game. Our results also show
that entanglement in the initial state can increase the payoff of
players compared with classical game.

This work is supported in part by China National Science
Foundation,
 the Fok Ying Tung
education foundation, the National Fundamental Research Program,
Contract No. 001CB309308 and the HangTian Science foundation.

\begin{figure}
\begin{center}
\includegraphics[width=8cm]{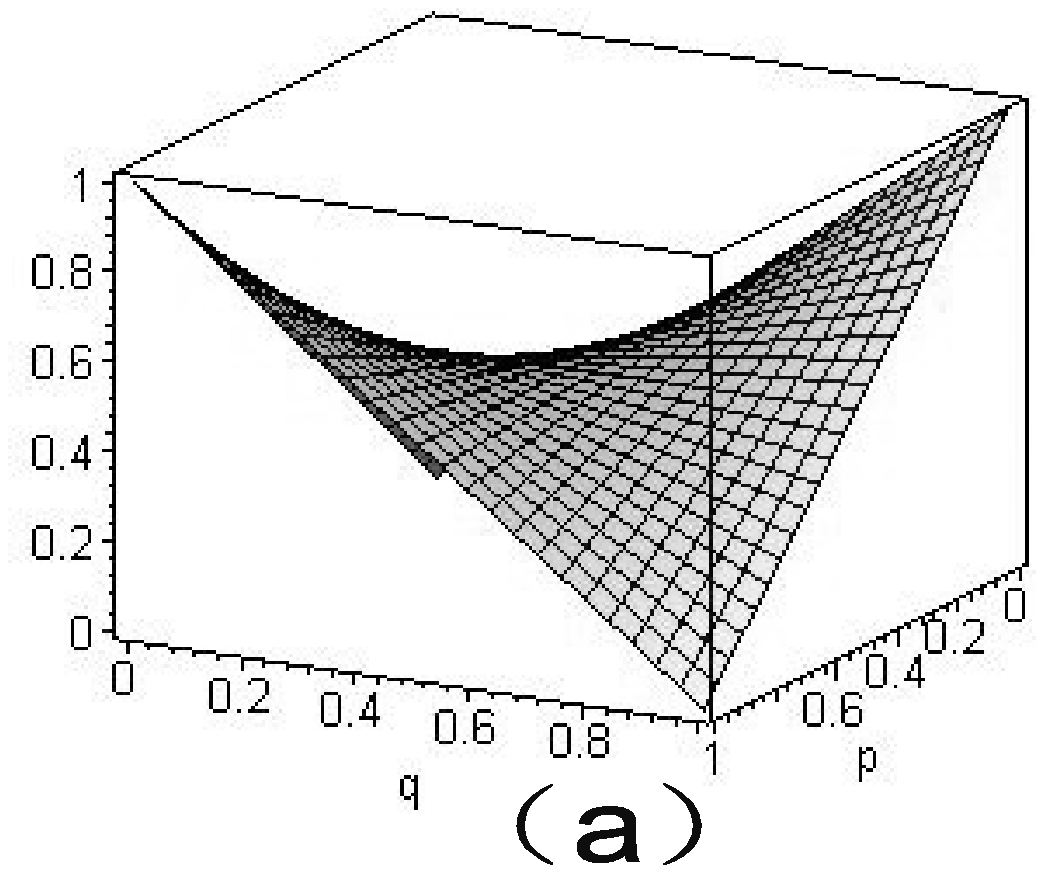}
\includegraphics[width=8cm]{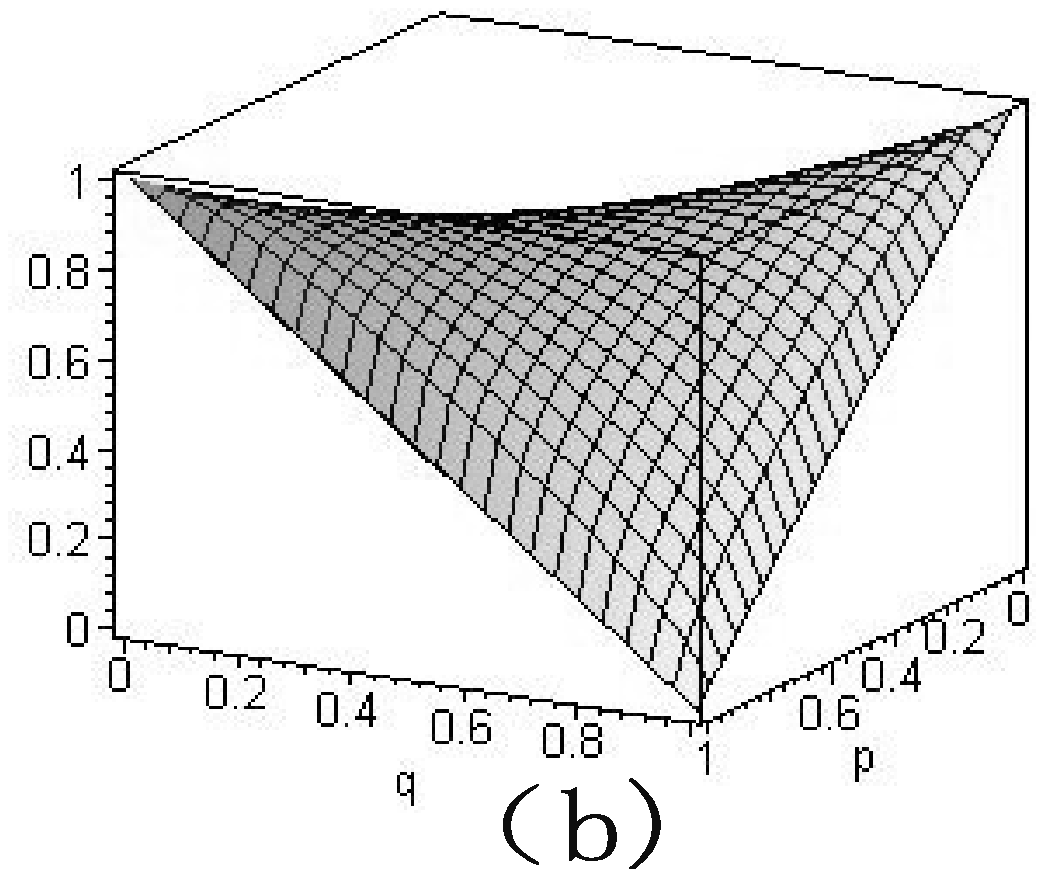}
\includegraphics[width=8cm]{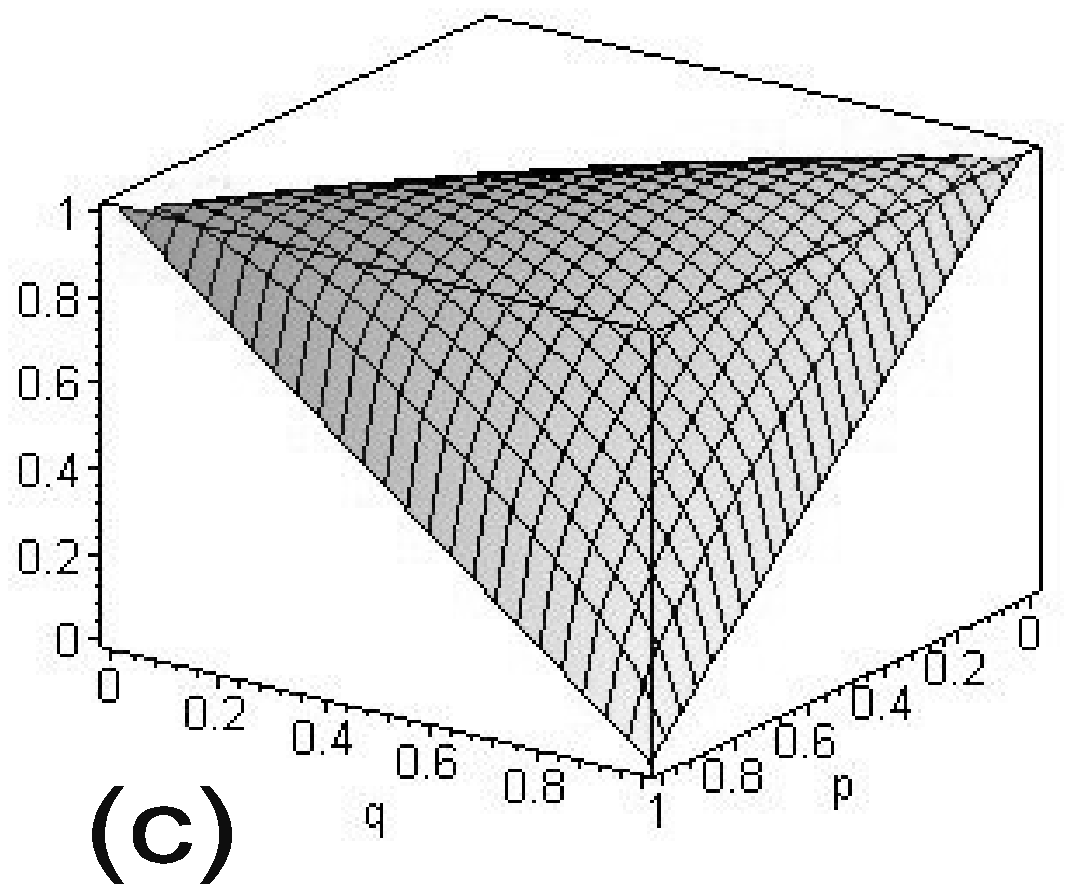}
\end{center}
\caption{Player A's payoff function versus $p$ and $q$ in
3-player cooperative via QSO strategy with initial state
$c_1|000\ket+c_2|111\ket$. Parameter $a$ takes the following
values: (a) $a=0$; (b)$a=0.33$; (c)$a=0.5$.}
\end{figure}

\newpage

\begin{figure}
\begin{center}
\includegraphics[width=8cm]{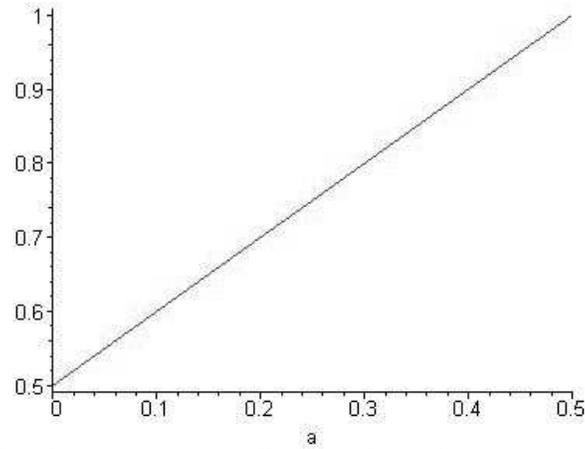}
\caption{The maximum payoff for player A versus coefficient $a$
for 3-player game with QSO strategy with initial state
$c_1|000\ket+c_2|111\ket$.}
\end{center}
\end{figure}

\begin{figure}
\begin{center}
\includegraphics[width=8cm]{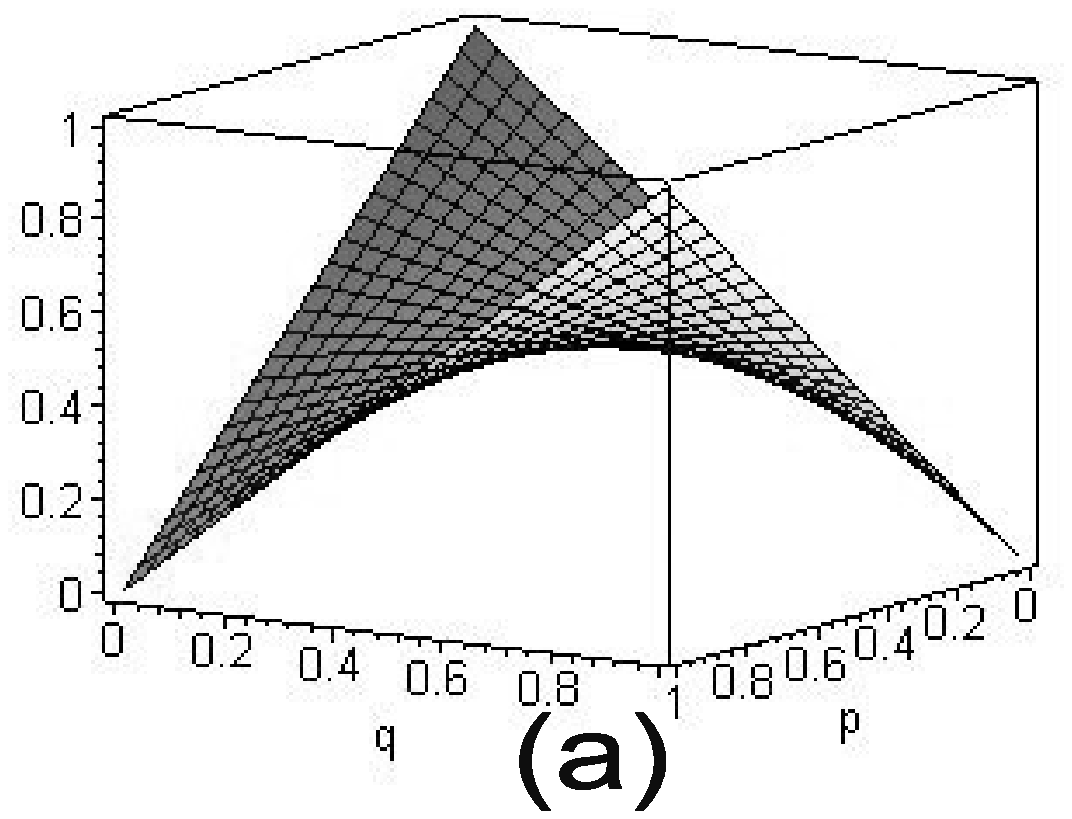}
\includegraphics[width=8cm]{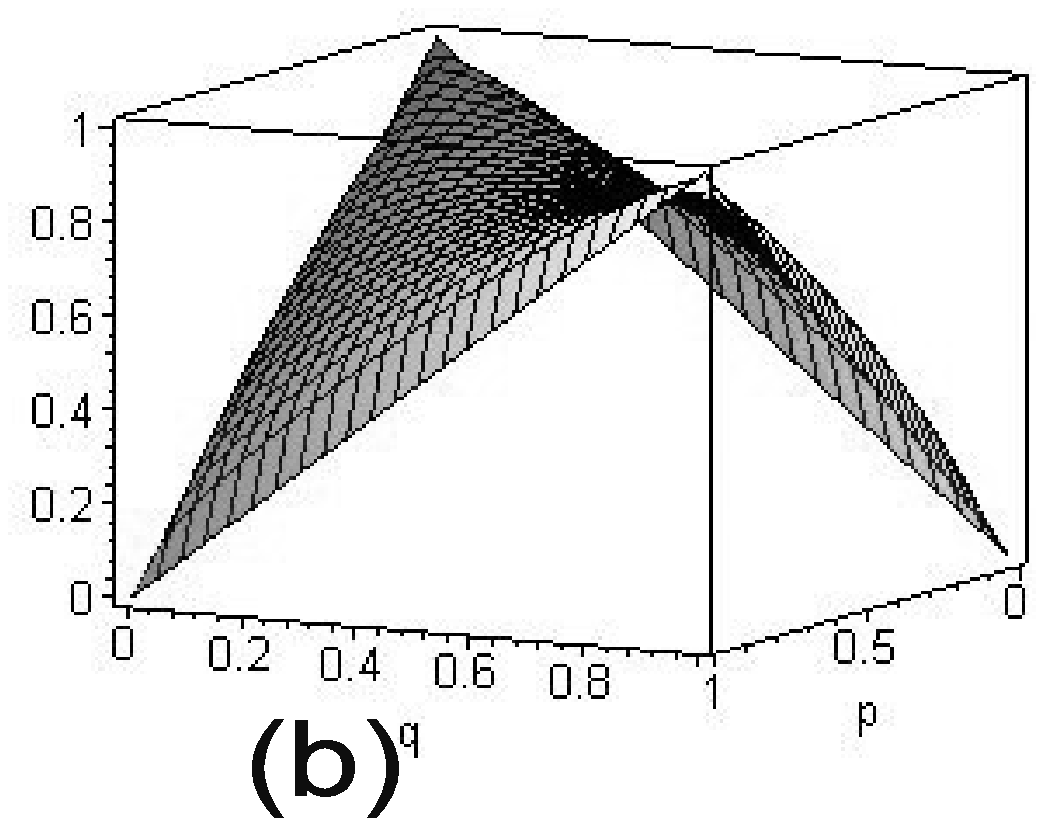}
\includegraphics[width=8cm]{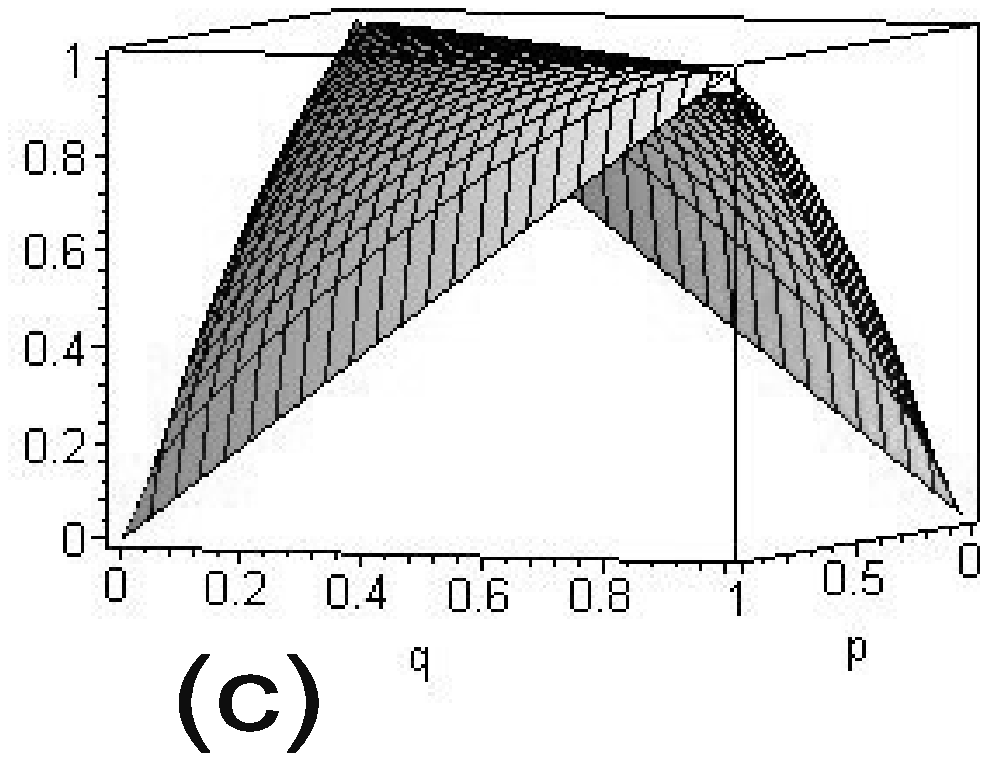}
\caption{Player A's payoff function versus $p$ and $q$ in
3-player cooperative via QSO strategy with initial state
$c_1|001\ket+c_2|110\ket$. Parameter a takes the following values:
(a) $a=0$; (b) $a=0.33$;(c)$a=0.5$. }
\end{center}
\end{figure}

\end{document}